\documentclass[12pt]{article}
\usepackage[toc,page]{appendix}
\usepackage{hyperref}
\usepackage{graphicx,pdflscape}
\usepackage{color}
\usepackage{bm}
\usepackage{alltt,dsfont}
\usepackage{appendix}
\usepackage{amsmath,amssymb}

\usepackage{tabularx}

\newcommand {\apgt} {\ {\raise-.5ex\hbox{$\buildrel\rangle\over\sim$}}\ }
\newcommand {\aplt} {\ {\raise-.5ex\hbox{$\buildrel\langle\over\sim$}}\ }

\newcommand{\iden}{ \mathds{ 1}}

\newcommand{\tr}{{\text {Tr} }}

\newcommand{\beq}{\begin{eqnarray}}
\newcommand{\eeq}{\end{eqnarray}}
\newcommand{\barray}{\begin{eqnarray}}
\newcommand{\earray}{\end{eqnarray}}
\newcommand{\nn}{\nonumber}

\usepackage{hyperref}
\hypersetup{
    colorlinks,
    citecolor=red,
    filecolor=cyan,
    linkcolor=blue,
   urlcolor=magenta
 }

\usepackage{xparse}
\ExplSyntaxOn
\NewDocumentCommand{\mref}{m}{\quinn_mref:n {#1}}
\seq_new:N \l_quinn_mref_seq
\cs_new:Npn \quinn_mref:n #1
 {
  \seq_set_split:Nnn \l_quinn_mref_seq { , } { #1 }
  \seq_pop_right:NN \l_quinn_mref_seq \l_tmpa_tl
  ( 
  \seq_map_inline:Nn \l_quinn_mref_seq
    { \ref{##1},\nobreakspace } 
  \exp_args:NV \ref \l_tmpa_tl 
  ) 
 }
\ExplSyntaxOff

\newcommand{\disp}[1]{Eq.~\mref{#1}}

\newcommand{\figdisp}[1]{Fig.~\mref{#1}}

\newcommand{\half}{\frac{1}{2}}

\usepackage{authblk}
\usepackage{amsthm}
\theoremstyle{plain}
\newtheorem*{theorem*}{Theorem}
\usepackage{graphicx} 
\graphicspath{{./Figures/}} 

\begin{document}
\title{  Ground-state  Entropy of the Ising model on a Frustrated lattice}
\author{ B Sriram Shastry$^{1}$\footnote{sriram@physics.ucsc.edu}, \;  
{Bill Sutherland$^{2}$\footnote{suther@mac.com}},\;
{Fr\'ed\'eric Mila$^{3}$\footnote{frederic.mila@epfl.ch}}\;  and 
{Afonso Rufino$^{3}$\footnote{afonso.dossantosrufino@epfl.ch}}\\ 
\small \em $^{1}$Physics Department, University of California, Santa Cruz, CA, 95064 \\
\small \em $^{2}$ 1789 Boxheart Dr,
Healdsburg, Ca 95448\\
\small \em $^{3}$ Institute of Physics
Ecole Polytechnique Federale de Lausanne,\\
\small \em BSP UNIL,
1015 Lausanne,
Switzerland}
\date{April 19, 2026}

\maketitle
\begin{abstract}
 We report  the ground-state entropy of a 2-d Ising model on the Shastry-Sutherland  lattice. We also study a generalization of this model, where a constraint on the zero temperature allowed configurations is removed  continuously.

\end{abstract}


\section{Introduction}
\begin{figure}[h]
\centering
\includegraphics[width=.45\columnwidth]{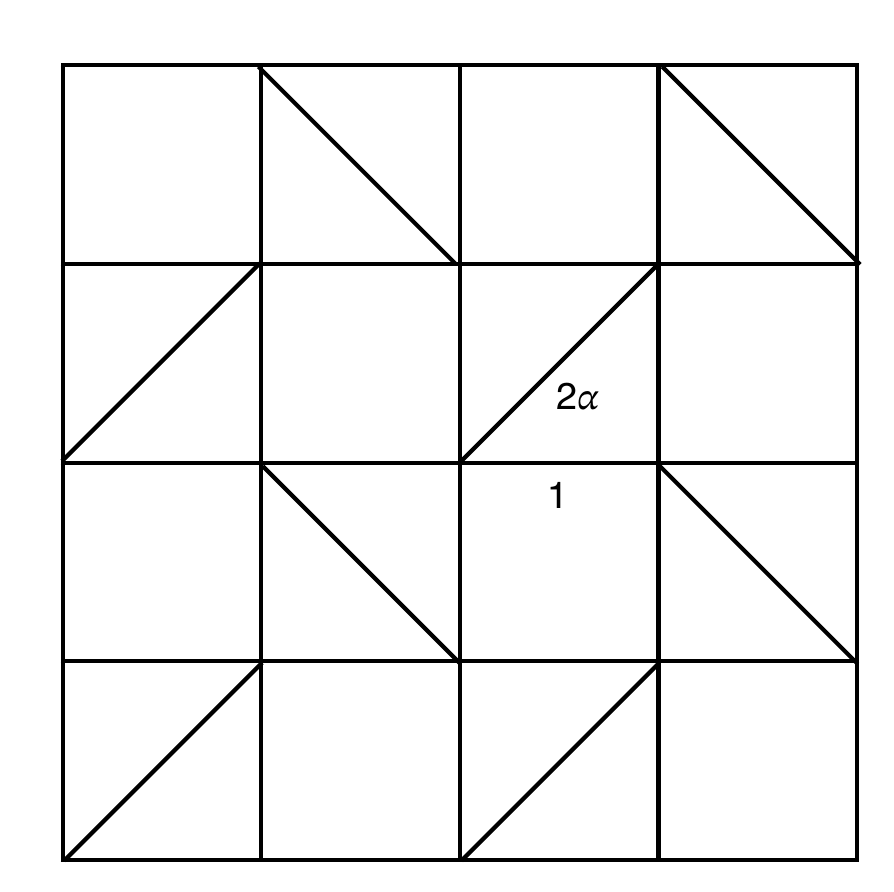}
 \caption{\footnotesize  The Ising model on the SS lattice. The weights in the Ising energy \disp{Ham-1} are $2 \alpha$  for diagonal bonds, and all others have weight 1. Stretching the bonds leads to other picturizations of this lattice as in \figdisp{Fig.2-3}  \label{Fig.1}}
 \end{figure}

\begin{figure}[h]
\centering
\includegraphics[width=.49\columnwidth]{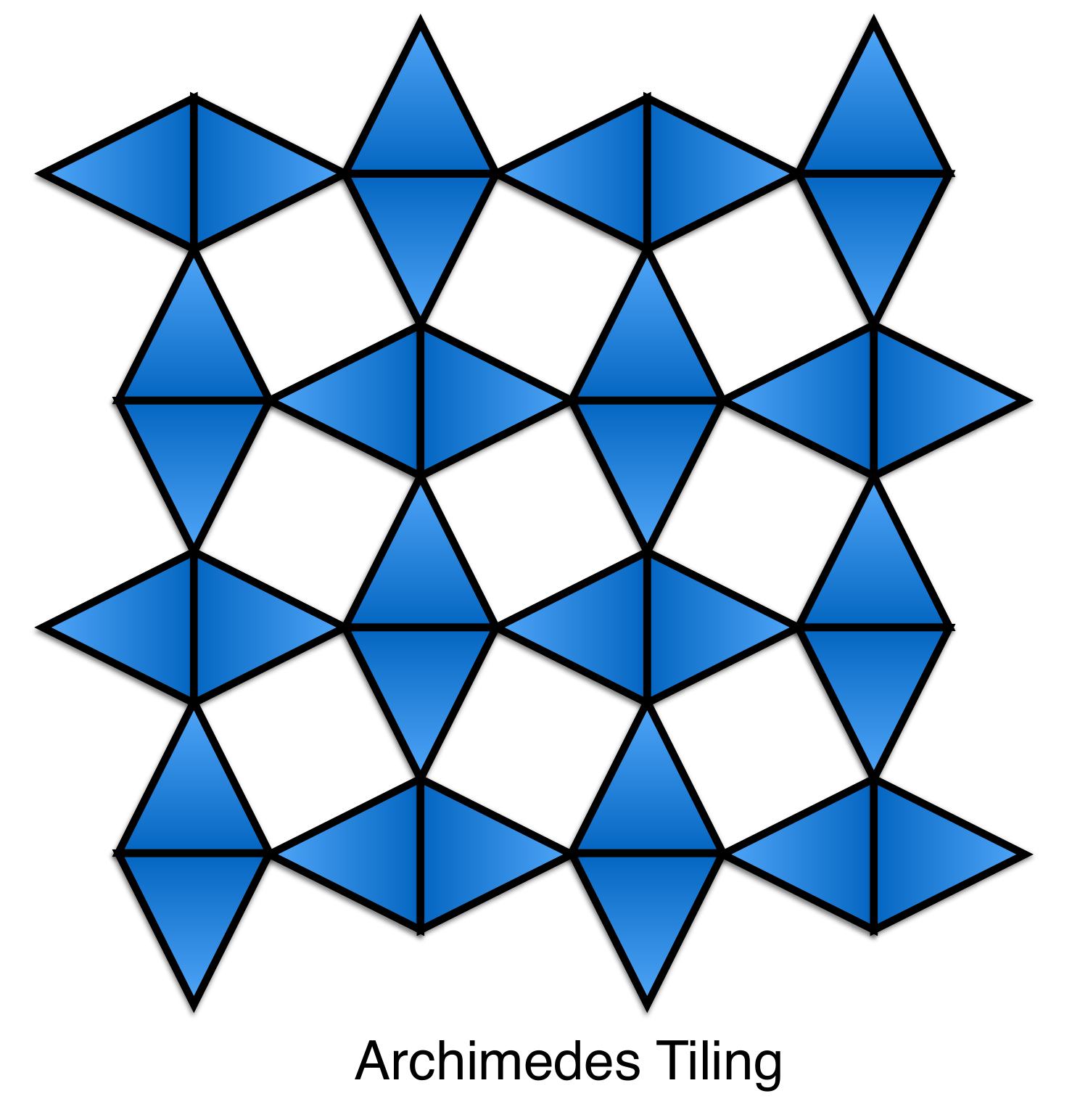}
\includegraphics[width=.49\columnwidth]{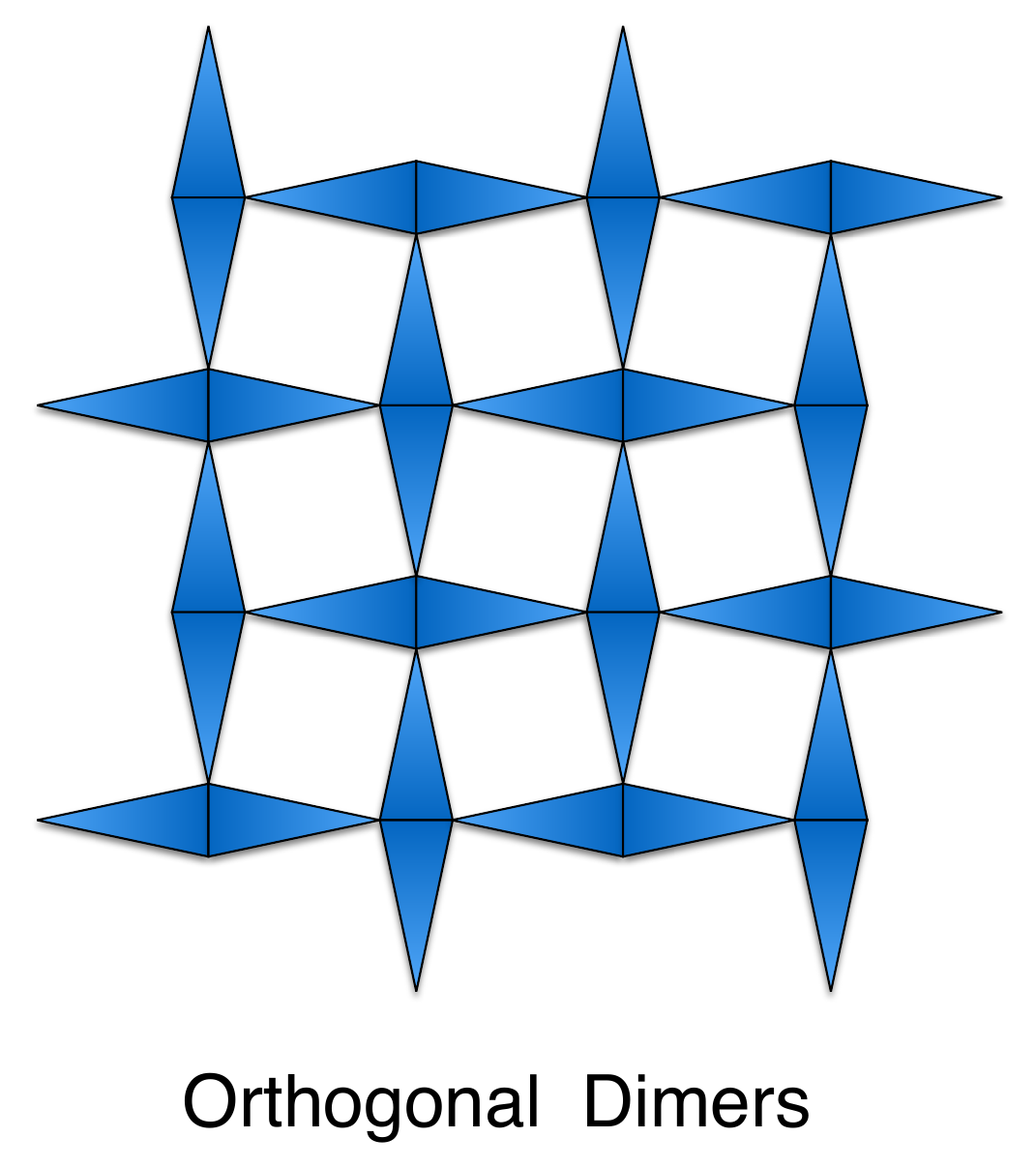}
 \caption{\footnotesize  Variants of the SS lattice. {\bf Left} Archimedean limit where equilateral triangle and squares with equal  bond length tile the 2-d plane. This is one of the 11 Archimedean lattices \cite{Grunbaum}.{\bf Right} The orthogonal dimer version where  the diagonal bonds of \figdisp{Fig.1} are shrunk relative to the sides of the squares. This picture is relevant to the measured lattice structure of $SrCu_2(BO_3)_2$ \cite{Keszler,Kageyama}. Since the diagonal bonds are relatively shorter than the other bonds, it  provides a  physical justification for $\alpha$$>$$1$, which is  necessary to describe its  magnetism.      \label{Fig.2-3}}
 \end{figure}

We  study the ground state entropy of the antiferromagnetic  Ising model on a 2-d frustrated lattice, referred to in literature as the Shastry-Sutherland (SS) lattice. This lattice and its variations related to it  by elastic deformations, the Archimedean lattice $(3^2,4,3,4)$, and the orthogonal dimer lattice, are illustrated in \figdisp{Fig.1,Fig.2-3}.  The quantum Heisenberg model on this lattice has attracted much interest in recent years \cite{Mila}. Interest peaked  after the realization that the magnetic system SCBO with a formula $SrCu_2(BO_3)_2$ provides an excellent physical  realization of the spin model\cite{Keszler,Kageyama,Ueda}.
It has led to considerable work on realizing a spin-liquid for suitable parameters of the model\cite{Kawakami,Sachdev}.

Our focus in this paper is on the antiferromagnetic  Ising model, where \cite{SS1} it was shown that the ground state is highly degenerate for  the parameter $\alpha$$\geq$$1$ (defined in \disp{Ham-1} and illustrated in \figdisp{Fig.1}). Two  lower bounds for the (extensive) entropy were given in \cite{SS1} showing its frustrated nature, analogous to that of the triangular lattice Ising model. However the exact entropy  remains unknown, and computing it is the goal for this work. We note that  the Ising model on the SSL has also received attention in different contexts  recently. Mappings to other lattices with known exact solutions  in certain parameter ranges  have been found. 
Stre\v{c}ka\cite{Strecka} used the star-triangle mapping of the Ising model on the SSL to a spin-$\half$ Ising model on the ``bathroom-tile'' or (4-8) lattice.  Rousochatzakis, L\"auchli and  Moessner \cite{Moessner} used a duality map of this model to the ``Cairo pentagonal lattice''. However the known exact solutions in these lattices are only available in regimes that exclude the problem of interest, namely the problem of ground state entropy of the SSL for the antiferromagnetic Ising model with $\alpha$$\geq$$1$.

 The Ising model is written as
\beq
H= \sum_{<\!ij\!>} \sigma_i \sigma_j + 2 \alpha \sum_{<lm>} \sigma_l \sigma_m  \label{Ham-1}
\eeq
where  $<\!ij\!>$ refer to bonds on the sides of squares, while   $<\!lm\!>$ refer to the bonds on the diagonals in  \figdisp{Fig.1}.

 At $\alpha=\half$, this model has a special symmetry since it has equal weightage to every bond in the lattice. In fact it  corresponds to a class of  lattices studied  by Archimedes in about 250 BC, who was interested in the geometric problem of tiling the 2-D plane with symmetric polygons. His musings brought forth  11 special lattices that bear his name. These Archimedean lattices were revisited by  Johannes Kepler, who discussed them  in his  panegyric  of symmetry in the physical world  titled   {\it Harmonice Mundi}\footnote{ Latin for  ``Harmony of the world''.} \cite{Kepler}, written in 1619.  By stretching the lattice  in \figdisp{Fig.1}  we get \figdisp{Fig.2-3}({\bf Left}). In this lattice, all bonds (edges) have the same length, and  each interior angle  in the triangles is  $\frac{\pi}{3}$. We may thus  view this lattice as consisting of (edge sharing) equilateral triangles and squares, all having equal sides. It then corresponds to tiling the 2-d plane with equilateral triangles and squares in a pattern  from Kepler's catalog\cite{Kepler}, termed as $(3^2,4,3,4)$ in  \cite{Grunbaum}. This name reflects the fact that going around {\em any}  lattice point, we  encounter a sequence of  2 triangles, a square, another triangle and finally another square.  This sequential arrangement is also visible in  \figdisp{Fig.1}, although the sides are not all equal and the symmetry is less than Archimedean. 

There is yet another representation, the orthogonal dimer picture shown in \figdisp{Fig.2-3}({\bf Right}), where the diagonal bonds are shortened relative to the other bonds. In this picture  the value of exchange interaction $\alpha$$>$$1$ is  intuitively reasonable, since within the theory of superexchange magnetism, increasing the bond strength follows from decreasing bond lengths. This picture is particularly  relevant to understanding the physics of materials like $SrCu_2(BO_3)_2$.


From a statistical mechanics point of view, varying the value of $\alpha$ is  of  interest. In \cite{SS1} the Ising model was shown to have 
 ground state entropy for $\alpha \geq 1$, while $\alpha$=$\half$ at the Archimedean limit,  is expected to have a non-degenerate ground state.  The value of the entropy is easily  computed  for $\alpha>1$, while for $\alpha=1$,  a  lower  bound was found in \cite{SS1} which establishes a degeneracy.  At $\alpha>1$, the degenerate ground state  manifold of states corresponds to spin pairs ($\uparrow$-$\downarrow$) or   ($\downarrow$-$\uparrow$) populating every $<\!lm\!>$ bond. Since the {\em number of such diagonal  bonds is $N/2$}, this twofold freedom yields a degeneracy  $2^{N/2}$, where $N$ is the number of sites.
 
 At  $\alpha$$=1$, a set of  new configurations come into play,
 these consist of parallel spins residing on the diagonal bond, which are missing at $\alpha>1$. We refer to the fraction of  these new  ferromagnetic diagonal type configurations as $n_{fmd}$, 
 \beq
 n_{fmd}= \frac{1}{N} \sum_{<l,m>} \langle {(1+\sigma_l \sigma_m)}\rangle  \label{nfmd},
 \eeq 
 where the average is over all ground-state configurations, and the sum is over all diagonal bonds as in \disp{Ham-1}. These new configurations arise in addition to all the ones contributing for $\alpha$$>$$1$, and therefore the entropy is certainly extensive. This implies that the entropy  at  $\alpha$$=1$ is bounded from below by that at $\alpha$$>1$.  In order to  characterize {\em all} of the ground state configurations,  we note that the total  energy \disp{Ham-1}  can be written exactly as the sum over energies of  triangles, each triangle has equal coupling and is of the form $(\sigma_a \sigma_b + \sigma_b \sigma_c+\sigma_c \sigma_a)$. Each diagonal bond supports two such triangles (whence the  requirement of $\alpha$$=$$1$ exactly), and the total number of triangles is therefore $N$.  

All the allowed  configurations in the ground state must satisfy the  condition that one bond  must be populated by a pair of antiparallel spins. The two remaining bonds must have one antiparallel pair and one parallel pair, due to the fundamental frustration of the triangle. Unlike the case  $\alpha$$>$$1$, the antiparallel pair of spins can populate any one of the three bonds in each triangle.   Enumerating all  configurations satisfying this condition accounts for  the degeneracy of the ground state.  We address this problem in the present work.

\section{Transfer matrix for calculating ground-state entropy}
\begin{figure}[t]
\centering
\includegraphics[width=.75\columnwidth]{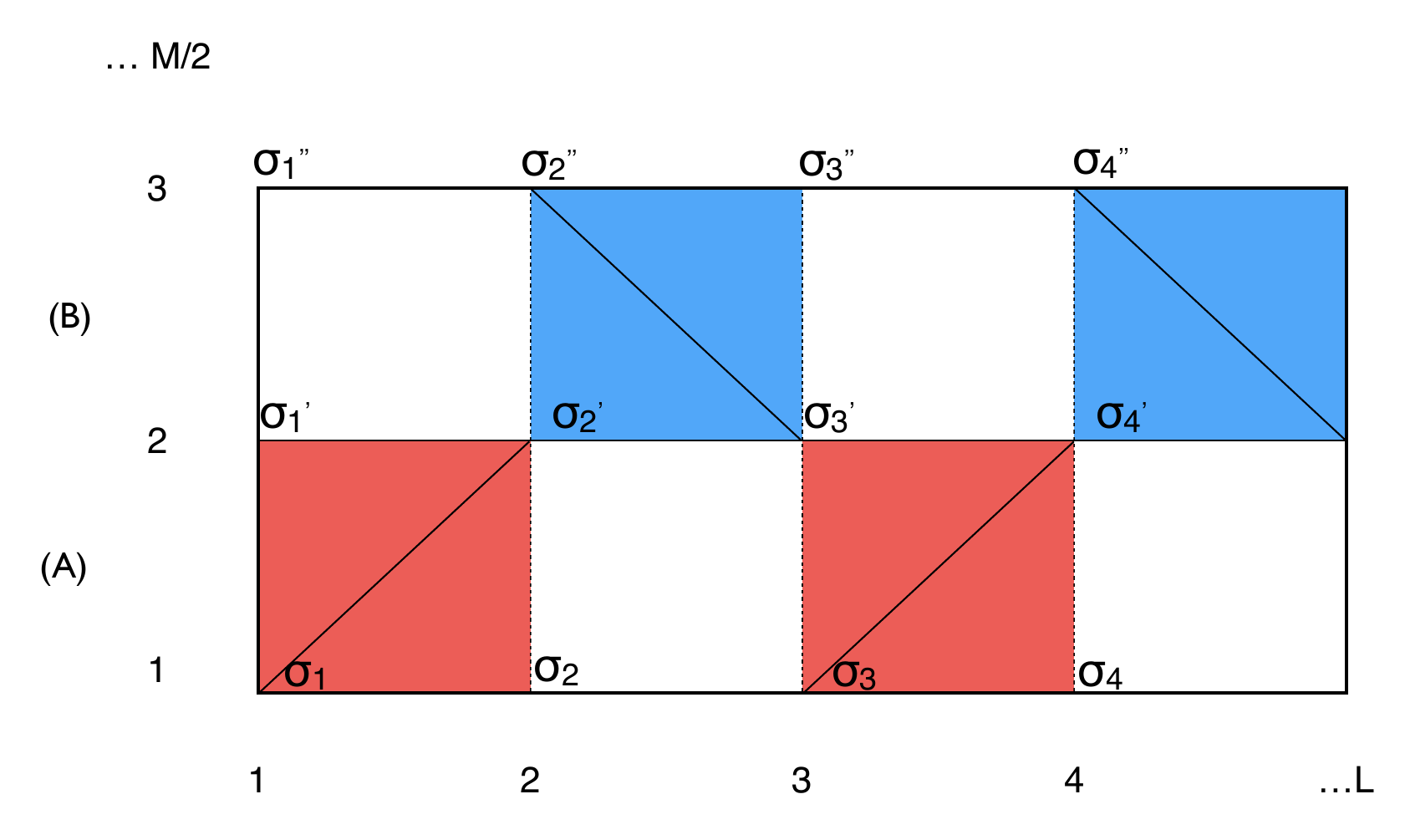}
 \caption{\footnotesize Two  rows of types (A) and (B) with diagonals slanting  distinctly. We represent the allowed  matrix elements in the notation $W^{(A)}(\sigma'_i,\sigma'_{i+1}|\sigma_i,\sigma_{i+1} )$ with odd $i$, and  $W^{(B)}$ for even $i$.    \label{Fig.4}}
 \end{figure}
 We now set up a transfer matrix for computing the entropy of the Ising model on the SS-lattice at T=0. It is also useful to consider a generalization where configurations with parallel spins on diagonal bonds, contributing to \disp{nfmd} are ``turned on'' continuously, with a parameter $r$. This parameter varies from $r=0$ without such configurations  to $r=1$ with the full complement of  such configurations\cite{Comment-1}.   This generalized model has a transfer matrix that is constructed below, the parameter ``r'' makes its appearance in  \disp{VA-1}, which gives the  Boltzmann weights $W$ in \disp{WA-1}.

  This formulation is  equivalent to the   arrow  vertex model    mentioned in  \cite{SS1}.  The SS-lattice is in many ways analogous to the checkerboard lattice introduced in \cite{Barma-Shastry}, for expressing Trotter formula decomposition of partition function of 1-d nearest neighbor quantum models such as the Heisenberg and  XYZ  models, in terms of the 2-d classical 6-vertex and 8-vertex models. If the diagonal bonds run in both directions, it would be exactly the same problem.

\begin{figure}[h]
\centering
\includegraphics[width=.75\columnwidth]{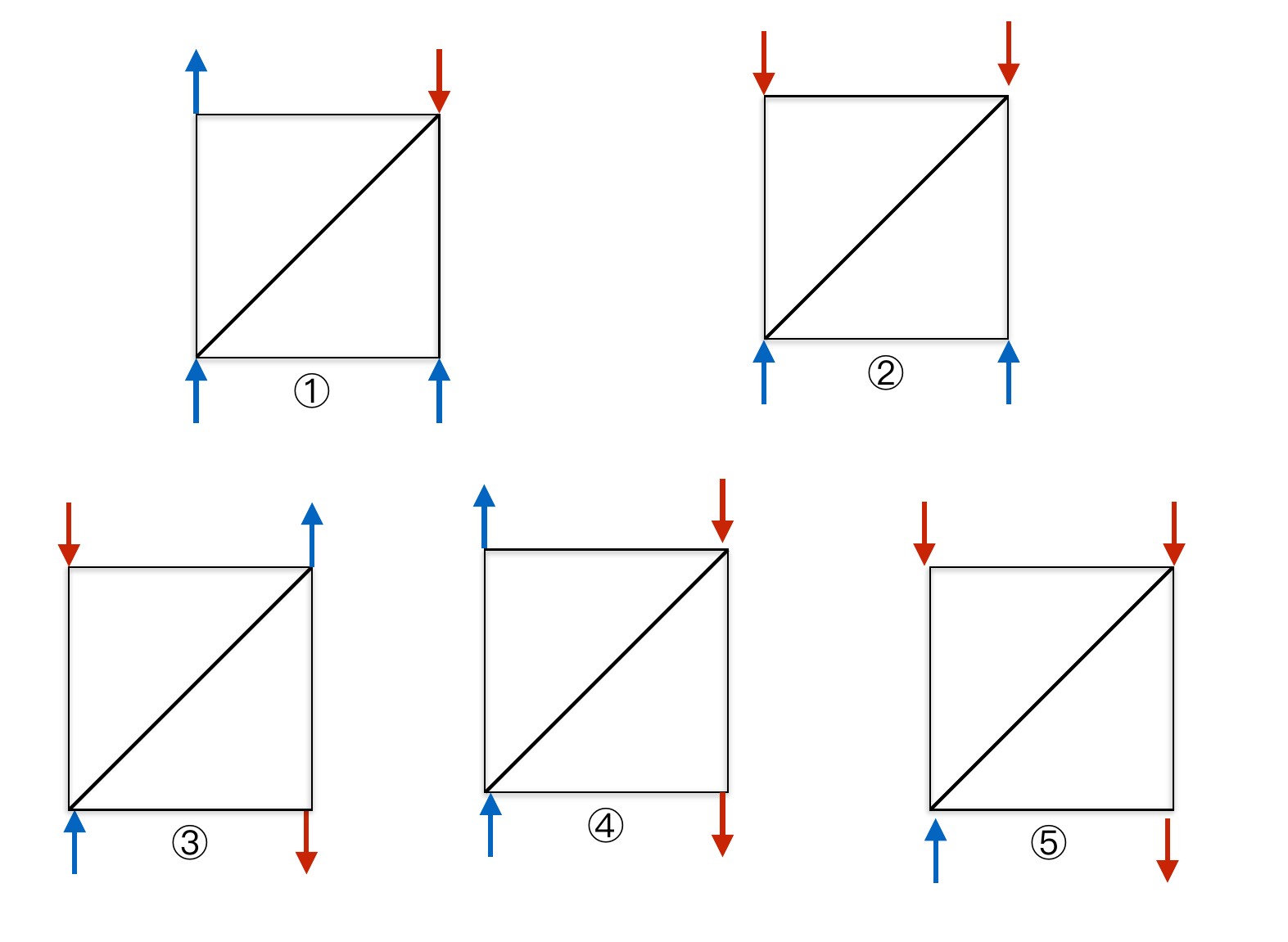}
 \caption{\footnotesize  The  configurations at each (A) type square $W^{(A)}(\sigma'_i,\sigma'_{i+1}|\sigma_i,\sigma_{i+1} )$, where $\sigma$ are the spins on the lower side and   $\sigma'$  are the  upper side.  The configuration \textcircled{3} is missing in the entropy calculation for $\alpha>1$, while all other configurations are common to the two problems.    A magnitude $r$ is assigned to  the diagram \textcircled{3}, and the rest are assigned 1. Varying $r$ from 0 to 1 gives a generalized model, where one can study the continuous  change in entropy between the two limits\cite{Comment-1}. Flipping initial and final state  spins leads  to allowed states with the same weight factors, and doubles the  allowed vertices, giving a total of  10 configurations.  \label{Fig.5}}
 \end{figure}

 Assuming periodic boundary conditions and even $L$
we  write the partition function (at T=0) for a lattice with sides $L\times M$ ($=N$)  in terms of a transfer matrix ${\cal T}_L$ for a chain of length $L$. For this purpose, following the conventions used in \figdisp{Fig.4}, we define the matrix elements  for rows of type (A) and (B) in terms of the non-zero matrix elements  $W^{(A)}(\sigma'_i,\sigma'_{i+1}|\sigma_i,\sigma_{i+1} )$ with odd $i$, and  $W^{(B)}$ for even $i$. The magnitudes of $W^{(A)}$ are assigned in \figdisp{Fig.5},
and we note the symmetry relation
\beq
W^{(B)}(\sigma'_b,\sigma'_a|\sigma_b,\sigma_a )=W^{(A)}(\sigma'_a,\sigma'_b|\sigma_a,\sigma_b ), \label{Wsymmetry}
\eeq
which stems from a mirror symmetry which flips the direction of the diagonals.

In terms of these, the two  transfer matrices ${\cal T}^{(A)}$ and ${\cal T}^{(B)}$ can be defined from their  matrix elements
 \beq \langle \{\sigma'\}|{\cal T}^{(A)}|\{\sigma\}\rangle=\prod_{odd \,i}W^{(A)}(\sigma'_i,\sigma'_{i+1}|\sigma_i,\sigma_{i+1} ) \label{TA-def-1}\eeq 
 and \beq\langle \{\sigma''\}|{\cal T}^{(B)}|\{\sigma'\}\rangle=\prod_{even \, i}W^{(B)}(\sigma''_i,\sigma''_{i+1}|\sigma'_i,\sigma'_{i+1} ).\label{TB-def-1} \eeq
It is convenient to transform these into an operator form using
\beq
\langle \sigma'_a,\sigma'_b|  V^{(A)}_{ab}|\sigma_a,\sigma_b \rangle =W^{(A)}(\sigma'_a,\sigma'_b|\sigma_a,\sigma_b ) \label{WA-1} \\
\langle \sigma'_a,\sigma'_b|  V^{(B)}_{ab}|\sigma_a,\sigma_b \rangle =W^{(B)}(\sigma'_a,\sigma'_b|\sigma_a,\sigma_b )\nn
\eeq
 The operator $V^{(A)}_{ab}$ is deduced by referring to \figdisp{Fig.5} as 
 \beq
V^{(A)}_{ab}&=& \frac{r}{2} \{ \sigma^x_a \sigma^x_b+\sigma^y_a \sigma^y_b \} +{\cal P}_{ab} \{ \sigma^x_a + \sigma^x_a  \sigma^x_b  \}\label{VA-1}\eeq
with a permutation operator  ${\cal P}_{ab}= \half(\iden+ \vec{\sigma}_a.\vec{\sigma}_b)$.
By  rearrangement, this can be written in the form
\beq
V^{(A)}_{ab}&=&\half \left[ \iden+ \sigma_a^x+\sigma_b^x+ i (\sigma_a^y \sigma_b^z-\sigma_a^z \sigma_b^y) -  \sigma_a^z\sigma_b^z + (r+1) \sigma_a^x\sigma_b^x + (r-1) \sigma_a^y\sigma_b^y\right]\nn \\ \label{VA-2}
\eeq
The parameter $r$   has been introduced in \disp{VA-1,VA-2} for studying the details of the transition from the case of  $\alpha>1$ (where $r=0$)  to  $\alpha=1$ (where $r=1$)\cite{Comment-1}.

We infer $ V^{(B)}$ using  the symmetry \disp{Wsymmetry} as
 \beq
 V^{(B)}_{ab}=V^{(A)}_{ba}. \label{VB}
 \eeq

With these definitions, we can write the partition function of  the Ising model on a $L\times M$  lattice as
 \beq
{\cal Z}_{LM}= \tr {\cal T}_L^{\frac{M}{2}} \label{Part-1}
\eeq 
where the trace $\tr$ is over the $2^L$ dimensional space of $L$ spin-$\frac{1}{2}$ particles. In the present case $\log {\cal Z}_{LM}$ counts the configurations at zero temperature.
 The exponent $\frac{M}{2}$  (rather than M) arises since each ${\cal T}_L$ contains  a {\em pair} of rows undergoing  scattering  as shown in \figdisp{Fig.4}. The transfer matrix is given by  
\beq
 {\cal T}_L &=& {\cal T}_B  \otimes {\cal T}_A  \label{Pair} \\
 {\cal T}_A &=& V^{(A)}_{L-1,L}\otimes V^{(A)}_{L-3,L-2} \ldots \otimes V^{(A)}_{1,2}   \label{TA} \\
 {\cal T}_B &=& V^{(B)}_{L-2,L-1} \otimes V^{(B)}_{L-4,L-3} \ldots \otimes V^{(B)}_{2,3},   \label{TB} 
 \eeq
 where ${\cal T}_A$ acts on initial states consisting of  spins $\{\sigma_j\}$ (the lowest row) and 
 ${\cal T}_B$ acts on initial states consisting of  spins $\{\sigma'_j\}$ (the upper row). The ${\cal T}'s$ follow from \disp{TA-def-1,TB-def-1}, and are   products of  scattering operators $V^{(A)}_{i, i+1}$ with odd $i$,  and $V^{(B)}_{i, i+1}$ with even $i$.

By the usual arguments, for large enough $L,M$,  we obtain  the entropy per site ${\bf \Sigma}$ (with $k_B=1$) from 
\beq
{\bf \Sigma} = \lim_{L,M \to \infty } \frac{1}{L M} \log {\cal Z}_{LM}. \label{entropy}
\eeq
If we denote the eigenvalues of ${\cal T}_L$ as $\lambda_j(L)$, the standard arguments obtain
\beq
{\bf \Sigma}=\lim_{L,M\to \infty } \frac{1}{L M} \log  \sum_j \{ \lambda_j(L)\}^ {\frac{M}{2}}.
\eeq
Further,  in most cases where $\lambda_j(L)$ has a maximum, say $\lambda^{max}$, that term dominates and we get
\beq
{\bf \Sigma}=\lim_{L\to \infty } \frac{1}{2 L } \log  \lambda^{max} (L) , \label{entropy-eigen}
\eeq

\subsection{\texorpdfstring{Results for entropy ${\bf \Sigma}$  at $\alpha=1$ from the Corner Transfer Matrix Renormalization Group}{Results for entropy Sigma at alpha=1 from the Corner Transfer Matrix Renormalization Group}}
Above, the calculation of the entropy of the SS lattice Ising model was formulated as an eigenvalue problem on the row-to-row transfer matrix $\mathcal{T}_L$. Even though this approach is exact in finite systems, its computational cost scales exponentially with system size and is hence limited to small system sizes.\par
\begin{figure}[h]
\centering
\includegraphics[width=\columnwidth]{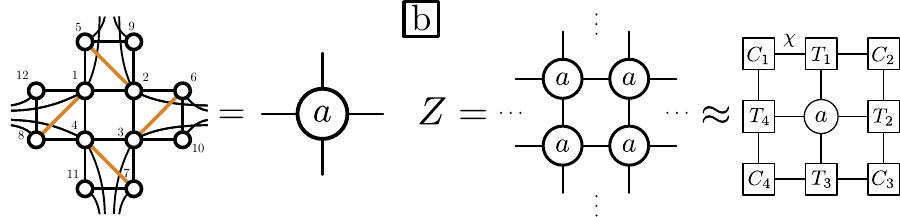}
 \caption{\footnotesize  Illustration of the CTMRG implementation of the SS lattice Ising model. (a) The partition function  of the system is written exactly as the contraction of a square network of fourth-order tensors $a_{i_1,i_2,i_3,i_4}$. (b) The contraction of all tensors surrounding a site is approximated by the product of finite corner $(C_i)$ and edge $(T_i)$ transfer matrices. }\label{fig:ctmrg}
 \end{figure}

An alternative approach is the Corner Transfer Matrix Renormalization Group (CTMRG) \cite{Baxter_CTM,Nishino_CTM}, a Tensor Network algorithm which generalizes the idea behind the Transfer Matrix to higher dimensions by expressing the partition function of a lattice system as the contraction of a grid of tensors. The computational cost of the exact contraction of a two-dimensional Tensor Network increases exponentially with system size, so in CTMRG one works directly in the thermodynamic limit and approximates the environment surrounding a site with corner $(C_i)$ and edge $(T_i)$ transfer matrices, which are truncated to a maximum dimension $\chi$ (see figure \ref{fig:ctmrg}(b) for an illustration of the method in diagrammatic notation \cite{Footnote-CTMRG}). $\chi$ controls the maximum correlation length that the method can describe, so in non-critical systems the exact thermodynamic limit is reached when the relevant observables are converged in $\chi$.
 In the specific case of the SS lattice  Ising model, the Tensor Network formulation of the partition function is obtained by letting each local tensor $a_{i_1,i_2,i_3,i_4}$ account for the Boltzmann weight of the 12-site cluster shown in figure \ref{fig:ctmrg}(a), with the indices imposing the matching conditions between contiguous clusters \cite{vanhecke}. 
 
Using CTMRG, we determined the entropy density ($\Sigma$), fraction of ferromagnetic diagonal bonds ($n_{fmd}$) and correlation length ($\xi$, in units of the lattice constant), shown in Table \ref{tab:ctmrg}. The results presented, obtained with $\chi=16$, are fully converged in $\chi$.
\begin{table}
\begin{center}
 \begin{tabular}{|c|c|c|}\multicolumn{3}{c}{\bf CTMRG results at $\alpha=1$}\\
  \hline
    $\Sigma$  & $n_{fmd}$ & $\xi$\\
    \hline
    0.45877772 & 0.20395387 & 0.476859\\
    \hline
  \end{tabular}
  \end{center}
  \caption{ { Entropy  ${\bf \Sigma}$ (\disp{entropy}),  the fraction of f.m. diagonals $n_{fmd}$ (\disp{nfmd})  and $\xi$ the spin-spin correlation length  $\alpha$=$1$ } obtained using CTMRG. All computed observables were found to be converged within machine precision for $\chi=16$.}
  \label{tab:ctmrg}
\end{table}


\subsection{\texorpdfstring{Results for entropy ${\bf \Sigma}$ from the transfer matrix.}{Results for entropy Sigma from the transfer matrix.}}
The eigenvalues of the transfer matrix ${\cal T}$  yield the entropy from \disp{entropy-eigen}, and while  it might be possible that these can be found analytically, we have not pursued it here. We can estimate the leading eigenvalue  by finite-size calculations for small $L$, and also variationally. 

{\bf \S  Largest Eigenvalue method Finite systems }
We calculated the largest eigenvalue of ${\cal T}$ using exact diagonalization of the transfer matrix for small systems. We used  the symbolic package {\em DiracQ} \cite{DiracQ} for this purpose.  The results for the entropy ${\bf \Sigma}$ are as follows:

\begin{table*}[h]
\begin{center}
\begin{tabular}{ |c|c|c|c|c||c| } \multicolumn{6}{c}{{\bf Entropy  per site ${\bf \Sigma}$,  of $L\times \infty$ system with $\alpha$=$1$ }}  \\ 
 \hline
 L & 4&6&8&10 & $L=\infty$\\ \hline
 ${\bf \Sigma}$&0.456592 &0.458905&0.458785&0.458775&0.45877772\\ \hline
 ${\bf \Sigma_{bound}}$&0.455578&0.456543&0.456678&0.456683& 0.45668258\\\hline
 \end{tabular}
\end{center}
\caption{\textcolor{black}{{\bf Second row}  First four columns show the calculated  entropy per site ($k_B=1$) at four values of L from the transfer matrix. The last column is from our numerical CTMRG study. It seems reasonable to infer from the finite $L$ data that, when  $L$$\to$$\infty$,    ${\bf \Sigma}$$\sim$$0.4588\pm0.0002$. This value is consistent with the exact CTMRG study at $L=\infty$. {\bf Third row}  first four columns show the computed lower bound to the entropy found from \disp{bound-entropy-eigen}, while the last column shows the infinite-size lower bound written in Eq.( \ref{eq:exact_lower_bound}). 
 \label{TabEntropy}}}
\end{table*}

{\bf \S  Variational Estimate of Largest Eigenvalue}

We can also find a lower bound to the entropy by replacing  the $\lambda^{max}(L)$ in  \disp{entropy-eigen} by the largest eigenvalue of ${\cal T}_L$ obtained within any finite dimensional subspace of states.  We make  Rayleigh-Ritz type ansatz, by taking the expectation in a  state $|\Psi_A\rangle$ which corresponds  to the largest eigenvalue of $ {\cal T}_A$, i.e. 
\beq
{\bf \Sigma}& \geq& {\bf \Sigma}_{bound}=\lim_{L\to \infty } \frac{1}{ 2 L } \log{ \langle\Psi_A | {\cal T}_B {\cal T}_A |\Psi_A\rangle}, \label{bound-entropy-eigen}
\eeq
so that 
\beq
 \langle\Psi_A | {\cal T}_B {\cal T}_A |\Psi_A\rangle&=&\lambda_A  \langle\Psi_A | {\cal T}_B  |\Psi_A\rangle,
\eeq
where $\lambda_A$ is found from a simple calculation
\beq
\lambda_A= \left( \frac{3+\sqrt{5}}{2}\right)^{\frac{L}{2}},
\eeq
and therefore the contribution from ${\cal T}_A$ to ${\bf \Sigma}_{bound}$  is  $\lim_{L\to \infty } \frac{1}{ 2 L } \log{ \lambda_A}=0.240606$ for any $L$. The contribution from the matrix element of ${\cal T}_B$ can be calculated numerically for finite $L$, as listed in Table~\ref{TabEntropy}.

\begin{figure}[t]
  \centering
  \includegraphics[width=\columnwidth]{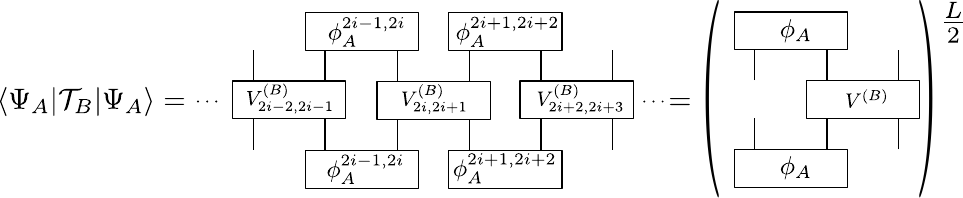}
  \caption{\footnotesize  Illustration of the exact calculation of the matrix element $\langle\Psi_A|\mathcal{T}_B|\Psi_A\rangle$ in Tensor Network diagrammatic notation \cite{Footnote-CTMRG}.}\label{fig:infinite_lower_bound}
\end{figure}

It is also possible to calculate $\langle \Psi_A|{\cal T}_B|\Psi_A\rangle$ exactly for infinite $L$, by exploiting the tensor product structure of $|\Psi_A\rangle$. Given the formula for $\mathcal{T}_A$ as a tensor product of two-site operators \eqref{TA}, the leading eigenvector can also be written as the tensor product
\begin{equation}
  |\Psi_A\rangle=\otimes_{i=1}^{L/2}|\phi_A^{2i-1,2i}\rangle,
\end{equation}
where 
{\footnotesize \begin{equation}
 |\phi_A^{2i-1,2i}\rangle =\frac{(1+\sqrt{5})(|\uparrow_{2i-1}\uparrow_{2i}\rangle+|\downarrow_{2i-1}\downarrow_{2i}\rangle)+(3+\sqrt{5})(|\uparrow_{2i-1}\downarrow_{2i}\rangle+|\downarrow_{2i-1}\uparrow_{2i}\rangle)}{\sqrt{40+16\sqrt{5}}}
\end{equation}}
is the largest eigenvector of $V^{(A)}_{2i-1,2i}$. In turn, since $\mathcal{T}_B$ is also a tensor product of two-site operators, as written in \eqref{TB}, it is possible to write $\langle \Psi_A|{\cal T}_B|\Psi_A\rangle$ as the tensor contraction illustrated in figure \ref{fig:infinite_lower_bound}. This tensor contraction equals the $L/2$'th power of a $4\times 4$ matrix which, in the $L\rightarrow\infty$ limit, is dominated by the largest eigenvalue:
\begin{equation}
  \langle \Psi_A|{\cal T}_B|\Psi_A\rangle\stackrel{L\rightarrow\infty}{=}\lambda_B^{L/2}=\left[\frac{1}{20} \left(4 \sqrt{5}+\sqrt{15 \left(8 \sqrt{5}+19\right)}+15\right)\right]^{L/2}.
\end{equation}
We therefore obtain the infinite-size lower-bound
\begin{equation}
  \begin{aligned}
    \Sigma_{Bound}&=\frac{\log(\lambda_A\lambda_B)}{4}=\frac{1}{4} \log \left(\frac{1}{40} \left(27 \sqrt{5}+\sqrt{3390 \sqrt{5}+7590}+65\right)\right)\\ 
    &=0.45668258\dots
  \end{aligned}\label{eq:exact_lower_bound}
\end{equation}

In \cite{SS1} the first two authors  used the same variational wavefunction   in a graphical framework, but due to a numerical error obtained   ${\bf \Sigma}_{Bound}=0.4812$, which is about $\sim$$ 5$\% larger than the exact value.

\section{Final Comments}

In this work we have studied the generalized Ising model on the SS-lattice. We note some results  that depend on the parameter ``r'', which plays the role of a conjugate field to the fraction of parallel spin pairs on diagonals $n_{fmd}$ (\disp{nfmd}). 

\begin{figure}[t]
\centering
\includegraphics[width=\linewidth]{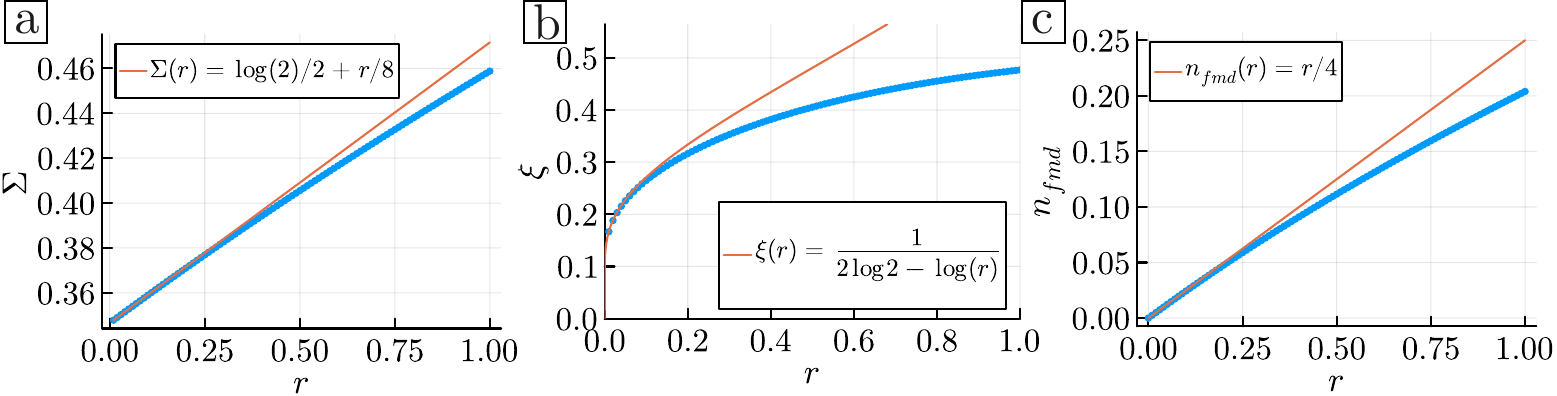}
 \caption{  \footnotesize  Entropy $(\Sigma)$, correlation length of spin-spin correlations $(\xi)$ and  fraction of ferromagnetic bonds ($n_{fmd}$) versus the parameter $r$ introduced  in \disp{VA-1,VA-2}, calculated with CTMRG. At $r=0$ we suppress configurations with parallel spins on the diagonals  (see  \figdisp{Fig.5}). This constraint  gives {\em all}  configurations for  $\alpha$$>$$1$, and by an elementary calculation gives ${\bf \Sigma}$=$\frac{1}{2}\log 2$. This is also the magnitude of the intercept at $r$=$0$ in the figure. At $\alpha$=$1$  we see from Table~\ref{TabEntropy} that ${\bf \Sigma}\sim 0.4589$, which matches the  figure.  In this figure we see a continuous increase of entropy as the constraint (of no parallel spins on diagonals) is relaxed continuously. In contrast, as a function of $\alpha$ the transition from $\alpha$=$1$ to $\alpha$$>$$1$ is abrupt.  The asymptotic formulas written in equations \eqref{eq:asymptotic_2} and \eqref{eq:asymptotic_1} are plotted in orange above the numerical results. \label{Fig-r}}
 \end{figure}

In figure \ref{Fig-r}, we display the variation of the entropy ${\bf \Sigma}$ with the parameter $r$ introduced in \figdisp{Fig.5} and \disp{VA-1,VA-2}.
We also show the evolution of $n_{fmd}$ with $r$, which shows a smooth evolution with $r$ from 0 to its maximum value, and of the correlation length $\xi$, which hosts a weak logarithmic singularity as $r\to 0$.

At the point $r$$=$$0$, the transfer matrix ${\cal T}$ (\disp{Pair}) has a particularly simple  flat eigenfunction,  where all configurations are added with equal weight 
\beq
|\Psi\rangle_0= \sum_{\sigma= \pm 1} |\{\sigma_j\}\rangle \label{special}
\eeq
 corresponding to the largest eigenvalue $2^L$ which leads to ${\bf \Sigma}$$=$$\half \log 2$. This curious fact is related to the fact that at $r$$=$$0$, for {\em any} pair of sites $a,b$, with operators defined in \disp{VA-1,VB},
\beq
V_{ab}^{(A)}|\Psi\rangle_0&=&2 |\Psi\rangle_0 \nn \\
V_{ab}^{(B)}|\Psi\rangle_0&=&2 |\Psi\rangle_0,
\eeq 
 i.e. share an eigenfunction despite the non-commutativity of these operators with sites that are common. This unusual situation, termed {\em superstability} in \cite{SS2},  points to the essential simplicity of the case of $\alpha$$>$$1$ (corresponding to $r=0$). For   $\alpha$$=$$1$  (corresponding to $r\neq0$), this superstability is lifted and the state corresponding to the maximum eigenvalue is non-trivial.

The {\em superstability} of the transfer-matrix also explains the weak singularity of the correlation length as $r\rightarrow 0$. Since $\left[\mathcal{T}(r=0)\right]^2 = 2^L\mathcal{T}(r=0)$, when $r=0$ the transfer matrix only has two distinct eigenvalues: the non-degenerate $\lambda^{max}(r=0)=2^{L}$ and the $(2^L-1)$-fold degenerate $\lambda^{(2)}(r=0)=0$. We determined the first-order perturbative correction to $\lambda^{max}$ explicitly,
\begin{equation}
  \lambda^{max}(r)\sim 2^{L}\left(1+\frac{rL}{4}\right),
\end{equation}
which leads to the asymptotic formulas
\begin{equation}
  \Sigma(r)\sim\frac{\log\lambda^{max}(r)}{2L}=\frac{\log 2}{2} +\frac{r}{8}, ~n_{fmd}(r)\sim 2r\frac{\partial\Sigma}{\partial r}=\frac{r}{4}.\label{eq:asymptotic_2}
\end{equation}
While in theory the correction to the second-largest eigenvalue $\lambda^{(2)}(r)$  can also be determined using degenerate perturbation theory, the very large degeneracy complicates this calculation. Instead, we rely on the CTMRG results, which show clearly that
\begin{equation}
  \lim_{r\to 0}\frac{\lambda^{(2)}(r)}{r \lambda^{max}(r)}=\frac{1}{4}\Rightarrow \lambda^{(2)}(r)\sim \frac{r}{4}2^{L}\left(1+\frac{rL}{4}\right).
\end{equation}
This allows us to propose the asymptotic formula for the correlation length
\begin{equation}
    \xi(r)=\left(\log\left|\frac{\lambda^{max}(r)}{\lambda^{(2)}(r)}\right|\right)^{-1}\sim\frac{1}{2 \log 2-\log r}\label{eq:asymptotic_1},
\end{equation}
which hosts a logarithmic singularity as $r\rightarrow 0$. The validity of the asymptotic formulas \eqref{eq:asymptotic_2} and \eqref{eq:asymptotic_1} can be confirmed by direct comparison with the CTMRG results in figure \ref{Fig-r}.


\begin{thebibliography}{99}


\bibitem{SS1}  B. S. Shastry and  B. Sutherland,  Exact Ground State of a Quantum-Mechanical Antiferromagnet,  Physica {\bf 108B}, 1069 (1981).

\bibitem{Mila} M. Albrecht and F. Mila, First-order transition between magnetic order and valence bond order in a 2D frustrated Heisenberg model, Europhys. Lett. 34 , 145 (1996).

\bibitem{Keszler} R. W. Smith and D. A. Keszler, Synthesis, structure, and properties of the orthoborate $SrCu_2(BO_3)_2$, J. Solid State Chem. {\bf 93}, 430 (1991)
\bibitem{Kageyama} H. Kageyama, K. Yoshimura,  N. V. Mushnikov, K. Onizuka, M. Kato, K. Kosuge,
C. P. Slichter, T. Goto, and Y. Ueda,  Exact Dimer Ground State and Quantized Magnetization Plateaus
in the Two-Dimensional Spin System $SrCu_2(BO_3)_2$,   Phys. Rev. Letts. {\bf 82}, 3168 (1999).
\bibitem{Ueda} S. Miyahara and K. Ueda, Exact Dimer Ground State of the Two Dimensional Heisenberg Spin System $SrCu_2(BO_3)_2$ , Phys. Rev. Letts. {\bf 82}, 3701 (1999).
\bibitem{Kawakami}  A. Koga and N. Kawakami,Quantum Phase Transitions in the Shastry-Sutherland Model for $SrCu_2(BO_3)_2$ , Phys. Rev. Lett. 84,  4461, (2000).
\bibitem{Sachdev}  C. Chung, J. Marston and S. Sachdev, Quantum phases of the Shastry-Sutherland antiferromagnet: Application to $SrCu_2(BO_3)_2$,  Phys. Rev. 64, 134407 (2001).


\bibitem{Strecka} J. Stre\v{c}ka, Exact solution of the spin-1/2 Ising model on
the Shastry-Sutherland (orthogonal-dimer)
lattice, Phys. Letts. {\bf 349},  505 (2006).

\bibitem{Moessner} I. Rousochatzakis, A. M. L\'auchli, and R. Moessner, Quantum magnetism on the Cairo pentagonal lattice, Phys. Rev. B {\bf 85}, 104415 (2012).


\bibitem{Kepler} Johannes Kepler, {\em Harmonice Mundi}, Linz (Austria): Johann Planck, (1619).

\bibitem{Grunbaum} Branko Gr\"unbaum and G. C. Shephard, {\em Tilings and Patterns},  Dover Publications, 2016. (ISBN-13978-0486469812).

\bibitem{Barma-Shastry} M. Barma and B. S. Shastry, Classical equivalents of one-dimensional quantum-mechanical systems, Phys. Rev. B {\bf 18}, 3351 (1978), see Fig. 1.

\bibitem{Baxter_CTM} Baxter, R. J., Corner transfer matrices. Physica A Statistical Mechanics and its Applications 106, 18–27 (1981).
\bibitem{Nishino_CTM} Nishino, T. and Okunishi, K., Corner Transfer Matrix Renormalization Group Method. J. Phys. Soc. Jpn. 65, 891–894 (1996).

\bibitem{vanhecke} Vanhecke, B., Colbois, J., Vanderstraeten, L., Verstraete, F. and Mila, F. Solving frustrated Ising models using tensor networks. Phys. Rev. Research 3, 013041 (2021).


\bibitem{Comment-1} In a finite T study of the Ising model, the  factor $r$  in the weight factor $W$ can be (somewhat artificially)  generated by adding  energy  $- \log r \times T$   for configurations with parallel spins on a diagonal. 

\bibitem{Footnote-CTMRG} In the Tensor Network diagrammatic notation, tensors are represented as squares or circles, while their indices are represented as lines. A line connecting two tensors stands for a contracted index. 

\bibitem{DiracQ} John G. Wright and B. S. Shastry, "DiracQ: A Quantum Many-Body Physics Package", arXiv:1301.4494[cond-mat.str-el] (2013), Journal of Open Research Software, 3(1) e13 (2015); DOI: https://doi.org/10.5334/jors.cb.

\bibitem{SS2} B. Sutherland and B. S. Shastry, `Exact Solution of a Large Class of Quantum Systems Exhibiting
Ground State Singularities, J. Stat.
Phys. {\bf  33}, 477 (1983).
\end{thebibliography}
\end{document}